%% file: manuscript.tex
\documentclass[pra,aps,showpacs,groupedaddress,superscriptaddress,twocolumn, numerical]{revtex4-1}

\input{./meta/packages}

\begin{document}
	
	\title{Precision Measurement of the $^{87}$Rb Tune-Out Wavelength \\
		   in the Hyperfine Ground State F=1 at 790 nm}
	\input{./meta/authors}
	\pacs{32.10.Dk, 37.10.Jk, 32.10.Fn}
	\date{\today}
	
	\begin{abstract}
		We report on a precision measurement of the $D$ line tune-out wavelength of $^{87}$Rubidium in the hyperfine ground state $\Ket{F=1, m_F=0,\pm1}$ manifold at $\SI{790}{\nm}$, where the scalar ac Stark shifts of the $D_1$ and the $D_2$ lines cancel. This wavelength is sensitive to usually neglected contributions from vector and tensor ac Stark shifts, transitions to higher principle quantum numbers, and core electrons. The ac Stark shift is probed by Kapitza-Dirac scattering of a Rubidium Bose-Einstein condensate in a one-dimensional optical lattice in free space and controlled magnetic environment.
		The tune-out wavelength of the magnetically insensitive $m_F=0$ state was determined to $\SI{790.01858 (23)}{\nano \meter}$ with sub ${\si{\pm}}$ accuracy. An \textit{in situ} absolute polarization, and magnetic background field measurement is performed by employing the ac vector Stark shift for the $m_F=\pm 1$ states. Comparing our findings to theory, we get quantitative insight into atomic physics beyond commonly used two-level atom approximations or the neglect of inner shell contributions.
	\end{abstract}	
	
	\maketitle
	
\section{Introduction}
	Energy shifts of atomic levels due to light-matter interaction have enabled optical traps for neutral atoms with numerous applications in state-of-the-art quantum technology as well as atomic and molecular physics. 
	Of particular interest are so-called magic wavelengths, where contributions originating from the coupling to different atomic transitions cancel in some quantity. 
	Important examples are the cancellation of differential light shifts in atomic traps for spectroscopic applications and metrology \cite{Barber2008, Akatsuka2008, Katori2011, Ushijima2015}, the minimization of differential light shifts for different hyperfine states \cite{Ludlow2008, Lemke2009}, or engineering of state-dependent traps \cite{Karski2009, Soltan-Panahi2011, Jimenez-Garcia2012, Belmechri2013}. 
		
	In mixed-species experiments, the usage of magic wavelength dipole traps facilitates engineering of species-selective optical traps, where in a mixture of two ultracold atomic species only one is optically trapped, while the other experiences a zero-crossing of the total energy shift for this so-called tune-out wavelength \cite{Leblanc2007, Arora2011}. Beyond the application of these tune-out wavelengths for quantum engineering, they yield information about the exact atomic level structure.
	This can be used to compare with \textit{ab-initio} calculations to refine the models of the atom and fundamental atomic data \cite{Arora2011}.	
	While there are versatile theoretical studies on tune-out wavelengths in alkali metals \cite{Leblanc2007, Rosenbusch2009, Arora2011, Safronova2012, Arora2012, Jiang2013, Jiang2013_2}, only few measurements have been performed in Potassium \cite{Holmgren2012, Trubko2015}, and Rubidium \cite{Herold2012, Leonard2015}, using interferometry and, respectively, light shift cancellation techniques.
	
	In our case, the system is designed to provide a species-selective optical lattice for single $^{133}$Cesium (Cs) atoms in a mixture with a $^{87}$Rubidium (Rb) Bose-Einstein condensate (BEC) \cite{Hohmann2015}. The tune-out wavelength for Rb at $\SI{790}{\nano\meter}$ results from blue detuning to the $D_1$ transition at $\SI{795}{\nano\meter}$ and red detuning to the $D_2$ transition at $\SI{780}{\nano\meter}$, so the ac Stark shifts cancel.
	
	Here, we report on the measurement of this tune-out wavelength in the ground state manifold $\Ket{F=1, m_F = 0,\pm1}$ in the absence of additional light fields and a controlled magnetic environment. Since the dominant scalar ac Stark shifts of both $D$ lines add to zero, our observation reveals usually neglected contributions, such as vector and tensor polarizabilities, transitions to higher quantum numbers, and the influence of core electrons to the scalar polarizability.

	We measure the tune-out wavelength of the $m_F=0$ state, where no vector ac Stark shift is present, with an accuracy improvement by a factor of $10$ compared to an earlier measurement \cite{Lamporesi2010}. We compare our value with a theoretical model, obtained from a recent measurement in a related system \cite{Leonard2015}.
	By employing the vector ac Stark shift in the case of $m_F=\pm1$, \textit{in situ} information about the absolute lattice polarization at a sub-percent level and the magnetic background field is gained.
	\section{Ac Stark shift and polarizability}
	To reveal the influence of additional contributions to the ac Stark shift, the scalar components from the $D$ lines have to be calculated accurately. Therefore, instead of using common approximations, \textit{i.e.} averaging the transitions' line widths \cite{Grimm2000} and neglecting the hyperfine structure (HFS) \cite{Leblanc2007}, we sum over all dipole allowed hyperfine transitions, coupled by the light field, following the formalism given in \cite{Rosenbusch2009, LeKien2013}.
	Here, the atom is interacting with an electromagnetic wave
	\begin{equation}
		\vec{E} = \frac{1}{2} E_0 \vec{\text{e}} e^{-i\left( \omega t - kz \right)} + \text{c.c.}
	\label{eq:runningWave}
	\end{equation}
	of amplitude $E_0$, frequency $\omega$, and wave number $k=\nicefrac{2\pi}{\lambda}$, propagating along the $z$ direction, where $\lambda$ is the laser wavelength. The vector $\vec{\text{e}}$ denotes an arbitrary complex polarization, that is described by a parametric angle $\theta_0$ as
	\begin{equation}
		\vec{\text{e}} = \hat{\text{e}}_x \cos{\theta_0} + i \hat{\text{e}}_y \sin{\theta_0},
	\label{eq:polarizationVector}
	\end{equation}
	with the degree of circular polarization $A = \sin{2 \theta_0}$.	
	The total energy shift is then calculated for an atom in the hyperfine Zeeman state $\Ket{ n \left( I J \right) F m_F }$, with main quantum number $n$, nuclear spin $I$, electronic angular momentum $J$, total angular momentum $F$ with $\vec{F} = \vec{J} + \vec{I}$, and its projection to the quantization axis $m_F$. This yields an ac Stark shift of
	\begin{equation}
	\begin{split}
		V^{\left(2\right)}_{n F m_F}(\lambda) = &- \left( \frac{1}{2} E_0 \right)^2 
		\Bigl[  \alpha^{\text{s}}_{nF}(\lambda)  + C \frac{m_F}{2F} \alpha^{\text{v}}_{nF}(\lambda)\\
		&- D \frac{3 m_F^2 - F \left( F + 1 \right)}{2F \left( 2F - 1 \right)} \alpha^{\text{T}}_{nF}(\lambda) \Bigr],
	\end{split}
	\label{eq:acStarkShift}
	\end{equation}
	with the state and wavelength dependent scalar, vector, and tensor polarizabilities $\alpha^{\text{s}}_{nF}(\lambda)$, $\alpha^{\text{v}}_{nF}(\lambda)$, and $\alpha^{\text{T}}_{nF}(\lambda)$ respectively, summing over all contributions from dipole allowed transitions to excited states $\Ket{ n^\prime \left( I J^\prime \right) F^\prime m_F^\prime }$. 
	The parameters $C=A \cos{\theta_k}$, and $D={\left( 3 \cos^2{\theta_p}-1 \right)}/{2}$ depend on the light field's polarization.
	Here, the orientation of the light field with respect to the system's quantization axis along the magnetic field vector $\hat{B}$ is given by the angle $\theta_k$ between the quantization axis and the wave vector, and $\theta_p$ between the quantization axis and the polarization vector, respectively.
	By factorizing the dependencies on $I$, $F$, $m_F$, $F^\prime$, and $m_F^\prime$, all polarizabilities are reduced to the dipole transition matrix elements $\Bra{n^\prime J^\prime} e \hat{r} \Ket{n J}$, which can be taken from literature.
	
	Note that only states with $m_F \neq 0$ show a vector ac Stark shift, which depends on the polarization of the light field.
	Nevertheless, the tensor polarizability, which vanishes for $J=\nicefrac{1}{2}$ if only fine structure splitting is taken into account \cite{Rosenbusch2009}, shows an ac Stark shift contribution even in the $m_F=0$ case.
	 
\begin{figure}[tb]
		\begin{center}
			\includegraphics[width=0.475\textwidth]{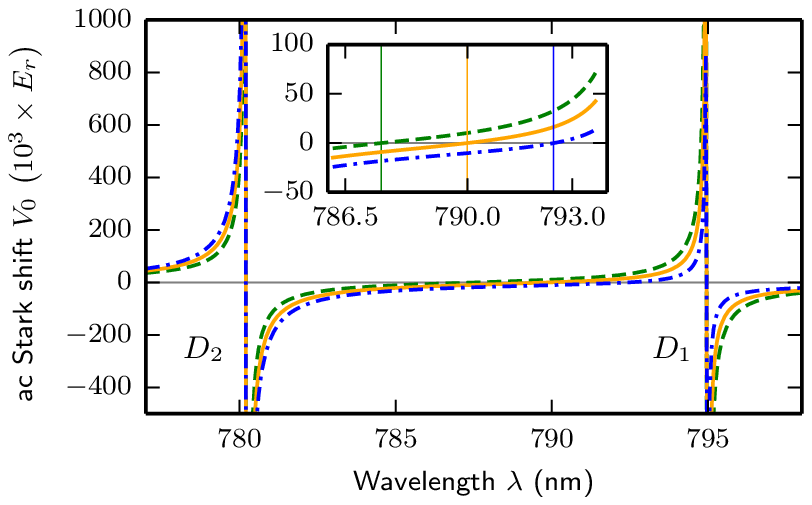}
		\end{center}
		\caption{The ac Stark shift $V_0$ for Rb atoms in the $m_F=0$ (solid), $m_F=+1$ (dashed), and $m_F=-1$ (dashed-dotted) ground states $\Ket{F=1}$, for right handed ($\sigma^+$) polarized light ($\theta_0 = \nicefrac{\pi}{4}$). 
		$V_0$ is given in multiples of the photon recoil $E_r$ at $\SI{790}{\nano \meter}$, and a typical light intensity in our experiment of $I=\SI{136}{\milli\W \per \centi\meter^2}$.
		The inset shows a magnified view of the tune-out wavelength for the three $m_F$ states (vertical lines). The tune-out wavelengths of the $m_F=\pm 1$ states are shifted by more than $\SI{2}{\nano \metre}$ with respect to the $m_F=0$ states due to the vector ac Stark shift.
		}
		\label{fig:DipolePotentialsPlot}
	\end{figure}

	Figure~\ref{fig:DipolePotentialsPlot} illustrates the wavelength and light polarization dependence of the ac Stark shift. The vector ac Stark shift disappears for $m_F=0$, but can shift the tune-out wavelengths of the $m_F=\pm1$ states by up to $\SI{2}{\nano \meter}$ in the case of circularly polarized light ($A=\pm1$).
	
\section{Experimental Method}
	We measure the ac Stark shift around the tune-out wavelength by employing Kapitza-Dirac (KD) scattering, the diffraction of a matter wave at a light grating. KD scattering has been originally introduced \cite{Kapitza1933} and measured \cite{Bucksbaum1988, Freimund2001} for electrons. The diffraction of neutral atoms was demonstrated for atomic beams \cite{Arimondo1979, Gould1986}, ultracold clouds \cite{Cahn1997}, and BECs \cite{Ovchinnikov1999}, and since then became a standard tool for atom interferometry applications \cite{Sapiro2009, Li2014} and optical lattice characterization \cite{Jo2012, Windpassinger2013, Cheiney2013}. \\
	We perform the KD experiment by flashing a Rb BEC for a duration $\tau=\SI{12}{\micro \second}$ with a one-dimensional static optical lattice, derived from two counterpropagating beams along $z$, with wavelength $\lambda$ and linear polarization parallel to $x$. The earlier discussed ac Stark shift results in a lattice potential of $V_0$.
	The Rb matter wave scatters at the light grating of periodicity $\nicefrac{\lambda}{2}$, imposing a momentum transfer of multiples of $2 \hbar k = 2\pi \hbar \frac{2}{\lambda}$, with $\hbar=\nicefrac{h}{2\pi}$ and Planck's constant $h$. In the theoretical description, the particle motion during the interaction is neglected (Raman-Nath regime) \cite{Kazantsev1980, Gould1986}, yielding an occupation of the momentum states $p = 2 N \hbar k$ ($N=0,\pm1, \pm 2, ...)$ with a respective probability of 
	\begin{equation}
		P_N( V_0, \tau ) = J^2_N\left( \frac{V_0 \tau}{2} \right),
	\label{eq:scatteringAmplitude}
	\end{equation}
	where $J_N(\theta)$ are the Bessel functions of the first kind. In our system, the Raman-Nath condition is fulfilled for absolute lattice depths $|V_0|\ll \SI{125}{E_r}$, given in multiples of the photon recoil $E_r = \nicefrac{\left(\hbar k\right)^2}{2m_{\text{Rb}}}$, with Rb mass $m_{\text{Rb}}$.
	\begin{figure}[tb]
		\begin{center}
			\includegraphics[width=0.48\textwidth]{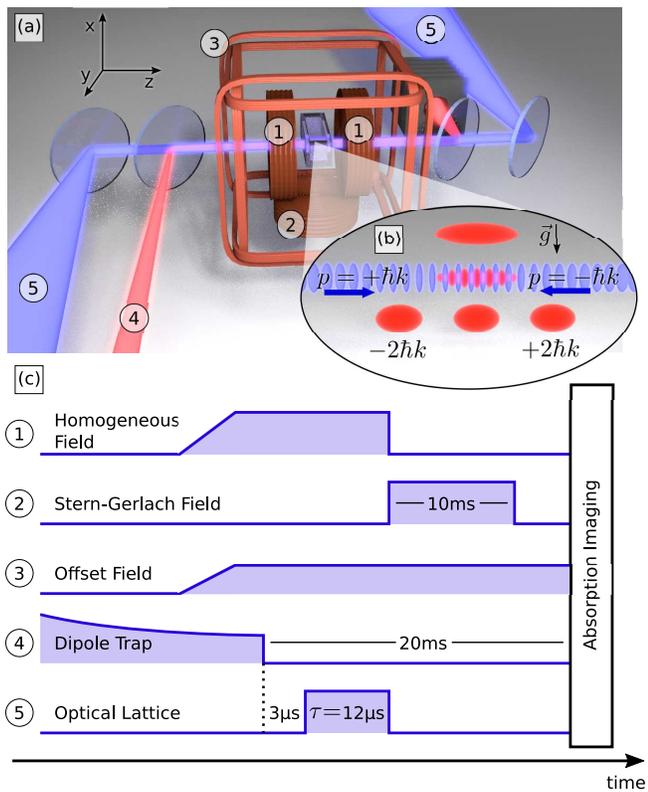}
		\end{center}
		\caption{Experimental Setup.
			(a) Simplified artist's view on the experiment. The coil setup used for magnetic field control consists of the homogeneous field coils (1), the Stern-Gerlach coil (2), and 3D compensation coils (3). The dipole trap beam (4) and the two counterpropagating lattice beams (5) are overlapped by dichroic mirrors on the main axis through the glass cell.
			(b) Schematic scattering process. When applying the lattice pulse on the free falling BEC, atoms are scattered into higher momentum orders $\pm 2 N \hbar k$ depending on the potential depth.
			(c) Schematic time line for the experimental sequence. After preparing the BEC in the dipole trap and setting the magnetic field to the desired value, the dipole trap is switched off and after a delay of $\SI{3}{\micro\second}$, the lattice pulse of $\tau=\SI{12}{\micro\second}$ duration is applied. Subsenquently, the Stern-Gerlach field gradient is switched on and after $\SI{20}{\milli\second}$ the atoms are imaged via absorption imaging along the y-axis.}
		\label{fig:Setup}
	\end{figure}

	Experimentally, both lattice beams are derived from a single-frequency Titanium Sapphire (Ti:Sa) laser and intensity-modulated with acousto-optic modulators (AOM) in each lattice arm, yielding a power of up to $\SI{450}{\milli \watt}$ per beam at a waist of $\SI{29}{\micro\meter}$. The AOMs shift the laser frequency by $-\SI{160}{\mega\Hz}\pm \SI{1}{\Hz}$ and therefore the wavelength by $\Delta\lambda=+\SI{0.42}{\pm}$ (at $\lambda=\SI{790}{\nano\meter}$).
	The radio frequency source, driving both AOMs is switched by means of a voltage controlled attenuator (VCA), limiting the pulse edges to a $\nicefrac{1}{e}$-time of $\SI{1.7}{\micro \second}$.
	In a crossed dipole trap at $\lambda=\SI{1064}{\nano \meter}$ a BEC of typically $2.5 \times 10^4$ atoms is prepared.
	The BEC is optically pumped to the absolute ground state $\Ket{F=1, m_F=+1}$ and driven into a desired $m_F$ state by means of a radio frequency transition.
	Details on our BEC preparation scheme are given in \cite{Hohmann2015}. 
	
	Although KD scattering experiments have also been performed in thermal gases \cite{Cahn1997}, the BEC system features a lower thermal momentum spread. 
	In particular, the thermal spread is smaller than the momentum transfer of multiples of $2 \hbar k$, which allows for separating and counting populations of higher momentum orders with standard time of flight (TOF) absorption imaging.
	
	Figure \ref{fig:Setup} shows our setup and a typical experimental sequence.
	The KD pulse is applied $\SI{3}{\micro \second}$ after releasing the BEC from the optical dipole trap, so we exclude any influence of the trap \cite{Neuzner2015} on the measured tune-out wavelength. 
	During the KD pulse, three orthogonal pairs of offset field coils create a magnetic field of up to $\SI{3}{G}$ in each direction.
	Additional homogeneous field coils allow to apply a strong offset field along $z$ of up to several $\SI{100}{G}$.
	Due to the VCA's switching response and fluctuations of the optical setup, the lattice pulse shape deviates from an ideal square wave. 
	Analyzing the time dependent intensity of both lattice beams, we obtain a reduction of the effective pulse duration from ideally $\SI{12}{\micro \second}$ to $\SI{8.75}{\micro \second}$
	with respect to the maximum lattice intensity, and an intensity fluctuation of $\SI{2.3}{\percent}$ rms, respectively.
	The latter results in a reduction of the lattice potential $V_0$ on the same order of magnitude. Note that the fluctuation of $V_0$ does not affect the zero-crossing point of the lattice potential.
	
\begin{figure}[tb]
		\begin{center}
			\includegraphics[width=0.5\textwidth]{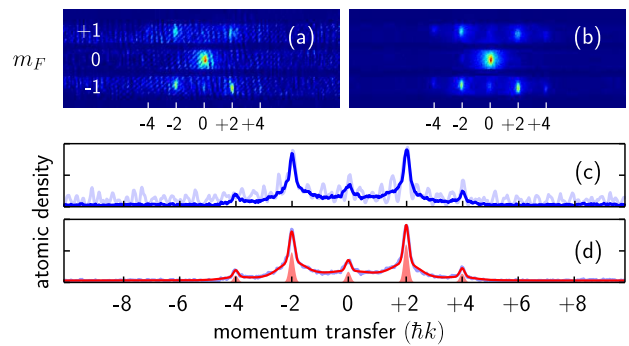}
		\end{center}
		\caption{KD image for a Rb spin mixture. (a) KD scattering image for $\SI{12}{\micro \second}$ pulse at a lattice wavelength of $\SI{790.018}{\nano \meter}$ after $\SI{20}{\milli \second}$ TOF. The $m_F$ states are distinguished by applying a magnetic Stern-Gerlach field along the gravitation axis $x$. The wavelength corresponds to the tune-out wavelength of the $m_F=0$ state. $m_F=+1$ and $m_F=-1$ show scattering into higher momentum orders due to the vector ac Stark shift. (b) The use of an improved absorption image calculation suppresses interference fringes of the imaging laser and background noise, increasing the SNR from $34$ to $95$.
		(c) The signal is vertically binned in the highlighted areas for each $m_F$ state. The improved image calculation (solid line) is compared to the standard technique (shaded). (d) A multi-Gaussian (solid line) is fitted to the line data (shaded line) from (b). The transfers are determined from the areas of the BEC peaks (shaded area). The population of the $\pm 4 \hbar k$ peaks correspond to less than $1750$ Rb atoms each.
		}
		\label{fig:AnalysisMethod}
	\end{figure}

	Figure \ref{fig:AnalysisMethod} shows a KD measurement of a Rb spin mixture after $\SI{20}{\milli \second}$ TOF. The magnetic states $m_F=0, \pm 1$ are spatially separated by applying an inhomogeneous magnetic Stern-Gerlach field after the KD flash.

	The population of each $ 2 N \hbar k$ momentum state is fitted with a double Gaussian, representing a superposition of a thermal background and a BEC peak. Since the thermal contributions cannot be well distinguished from each other, only the BEC contributions are included into the analysis according to equation  (\ref{eq:scatteringAmplitude}). Note that the scattering amplitudes $P_N(V_0, \tau)$ are equal for positive and negative lattice potentials, and therefore give a measure of the absolute ac Stark shift. Therefore, the sign has to be determined from the theoretical model.
	
	When measuring KD at weak lattice potentials $V_0$, the atomic density signal in higher momentum states approaches the detection threshold of our imaging system, which is mainly determined by the occurrence of fringes in the absorption images.
	Since the atomic density is calculated as $\propto -\log{(S_i / R_i)}$ from the actual absorption image $S_i$ containing the shadow of the BEC, and a reference image $R_i$ with only the probe laser light, any fluctuation of interference patterns between the two images results in fringes in the atomic density (see figure \ref{fig:AnalysisMethod}(a)).
	
	This effect is strongly suppressed by using an optimum reference image $R_{\text{best}}$, that reproduces the background properties of the signal shot $S_i$~\cite{Ockeloen2010}, and thereby the specific interference pattern.
	We compute $R_{\text{best}}$ as a linear combination $\sum_k^M c_k R_k$ of up to $500$ reference images from a base $\left[ R_k \right]$, so it optimally matches a signal-free area in the absorption signal $S_i$ with a size of roughly $200 \times 300$ pixel.
	The least square fitting algorithm, given in \cite{Ockeloen2010} was optimized by diagonalizing the base of reference images $\left[ R_k \right]$, allowing for a computation time of less than $\SI{100}{\milli \second}$
	per absorption image on a standard personal computer.
	The approach avoids fringes, created by time varying interference patterns of the probe laser beam at the best and even yields a reduction of photon shot noise ~\cite{Ockeloen2010}, increasing the signal to noise ratio (SNR) by a factor of at least $3$ in our measurements.
\section{Results}
\subsection{$m_F=0$ State Measurement}
	\begin{figure}[htb]
		\begin{center}
			\includegraphics[width=0.5\textwidth]{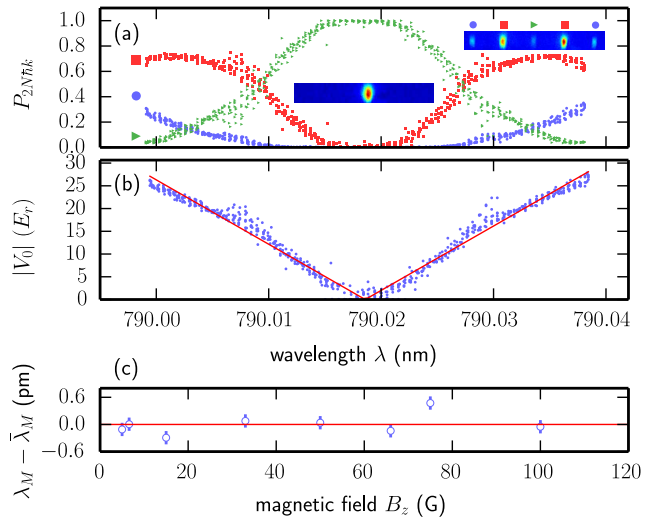}
		\end{center}
		\caption{Tune-out wavelength measurement of Rb in the $\Ket{F=1, m_F=0}$ state. (a) From the KD scattering images, the populations in the zero momentum state (triangle), $\pm 2 \hbar k$ (square), and $\pm 4 \hbar k$ (circle) state were extracted. Insets show samples of KD images for vanishing (center) and maximal potential (right). (b) The resulting lattice potential $V_0$ (dots) was fitted with the model of eq.~(\ref{eq:linearApprox}) with the tune-out wavelength $\lambda_M$ (solid line). (c)~For various magnetic offset fields $B_z$ the tune-out wavelength has been measured. The values scatter around the average of $\bar{\lambda}_M = \SI{790.01858}{\nano \metre}$ (solid line). The error bars are calculated from the fitting uncertainties of  (b) and the wavelength measurement calibration.}
		\label{fig:mFMagicWavelength}
	\end{figure}
	We measure the tune-out wavelength of the $m_F = 0$ state, where the scalar and the tensor, but no vector ac Stark shift is present.
	The laser wavelength is measured with a wavelength meter, offering a resolution of $\SI{50}{\mega\hertz}$,
	and an automatic calibration to a built-in wavelength standard. We have validated the calibration by measuring the wavelength of a laser, stabilized to the Rb $\Ket{5S_{\nicefrac{1}{2}}F=1}\rightarrow\Ket{5P_{\nicefrac{3}{2}}F=2}$ transition \cite{Steck2010}, with an absolute accuracy of $\SI{20}{\mega\hertz}$.
	For the measurements we use a magnetic offset field of $(B_{x_0}, B_{y_0}, B_{z_0}) = (0.4, 0.3, -0.7) \si{G}$ and apply a homogeneous quantization field $B_z$ along $z$. Since for $m_F=0$ the nonlinear Zeeman energy
	is negligible for weak magnetic fields, we do not expect a magnetic field dependence of the tune-out wavelength.	
	However, the measurement is limited to $B_z\approx \SI{100}{G}$. For higher fields, an asymmetric background in the absorption images occurs, which perturbs the momentum transfer fits. We attribute the asymmetric background to a superposition of the homogeneous quantization field and the Stern-Gerlach field.

	\setlength{\extrarowheight}{.5em}
	\begin{table*}[htb]
		\caption{Comparison of the measured tune-out wavelength $\lambda_{M, \text{experiment}}$ of the Rb $\Ket{F=1, m_F=0}$ state with a theoretical model. First, the tune-out wavelength $\lambda_{M, \text{5s-5p, theo}}$ is calculated from the wavelength dependent $D$ polarizability. 
		We compare predictions, using the reduced dipole matrix elements from direct measurement (a) from \cite{Steck2010}, and the more accurate reduced dipole moment $d_{\nicefrac{3}{2}}$ from \cite{Steck2010}, and the ratio $R = \nicefrac{|d_{\nicefrac{3}{2}}|^2}{|d_{\nicefrac{1}{2}}|^2}$ from \cite{Leonard2015} (b), respectively.
		A more accurate model $\lambda_{M, \text{total, theo}}$ is obtained, when further polarizability contributions are included. Each shifts the tune-out wavelength by $\Delta\lambda_{\text{tensor}}$ (tensor) $\Delta\lambda_{\text{5s-6p+}}$, (transitions to higher quantum numbers), and $\Delta\lambda_{\text{c, cv}}$ (core electrons, core-valence electron interaction). Polarizabilities of the latter two components are taken from \cite{Leonard2015}.} 
		\begin{tabularx}{17.8cm}{>{\centering\arraybackslash}XXXXXXX}
			\hhline{=======}
			\multirow{2}{*}{matrix elements}&  $\lambda_{M, \text{5s-5p, theo}}$\quad $+$& $\Delta\lambda_{\text{tensor}}$\quad $+$ & $\Delta\lambda_{\text{5s-6p+}}$ \quad $+$& $\Delta\lambda_{\text{c, cv}}$ \quad $=$& $\lambda_{M, \text{total, theo}}$ & 
			{$\lambda_{M, \text{experiment}}$} \\
			
			& $\left( \si{\nano \meter} \right)$ & $\left( \si{\pico \meter} \right)$ & $\left( \si{\pico \meter} \right)$ & $\left( \si{\pico \meter} \right)$ & $\left( \si{\nano \meter} \right)$ & $\left( \si{\nano \meter} \right)$\\
			\hline
			
			$d_{\nicefrac{1}{2}}$, $d_{\nicefrac{3}{2}}$$^{\text{(a)}}$ & $790.0181(56)$ & \multirow{2}{*}{ $0.091(1)$} &  \multirow{2}{*}{$1.203(48)$} & \multirow{2}{*}{$3.455(38)$} & $790.0228(57)$ & \multirow{2}{*}{$790.01858(23)$}\\ 
			
			$R$, $d_{\nicefrac{3}{2}}$$^{\text{(b)}}$ & $790.01374(1)$ &  &  & & 790.01850(9) \\
			\hhline{=======}					
		\end{tabularx}
		\label{tab:TheoryMagicWavelength}
	\end{table*}
	Figure \ref{fig:mFMagicWavelength} shows the measurement of the lattice potential depth around the tune-out wavelength for various offset fields $B_z$ in a range of $\SI{5}{G}$ to $\SI{100}{G}$. The dependence of the potential $V_0(\lambda)$, taken from equation (\ref{eq:acStarkShift}) is approximately linear with a maximum deviation of $\SI{0.25}{\percent}$ in the measured wavelength range, and writes
	\begin{equation}
		V_0(\lambda)= \frac{\partial V_0}{\partial \lambda} (\lambda-\lambda_M),
		\label{eq:linearApprox}
	\end{equation}
	with the slope $\frac{\partial V_0}{\partial \lambda}$, and the tune-out wavelength $\lambda_M$. 
	Since the KD scattering analysis yields absolute values, $\left|V_0(\lambda)\right|$ is fitted to the potentials. 
	We average the $\lambda_M$ data and obtain a tune-out wavelength of $\bar{\lambda}_M = \SI{790.01858 (23)}{\nano \meter}$
	for the $\Ket{F=1, m_F=0}$ state, providing a 10-fold accuracy improvement compared to the previous measurement of $\SI{790.018(2)}{\nano \meter}$ in the same internal state \cite{Lamporesi2010}.
	We compare our result to the model from equation (\ref{eq:acStarkShift}), assuming a scalar polarizability of
	\begin{equation}
		\alpha^{\text{s}}_{nJF}(\lambda) = \alpha^{\text{5s-5p}}_{F}(\lambda) + \alpha^{\text{5s-6p+}}_{J} + \alpha^{\text{c, cv}}.
	\end{equation}
	Here, $\alpha^{\text{5s-5p}}_{F}(\lambda)$ contains all wavelength dependent scalar polarizabilities from the Rb $D$ lines, that we calculate from the dipole allowed hyperfine transitions with respective energy splitting \cite{LeKien2013}.
	Transitions to higher $P$ states are represented by $\alpha^{\text{5s-6p+}}_{J}$, where only the fine structure is taken into account. The contribution of core electrons and core electron - valence electron interaction is represented by $\alpha^{\text{c, cv}}$ \cite{Arora2011}. For $\alpha^{\text{5s-6p+}}_{J}$ as well as $\alpha^{\text{c, cv}}$ most recent values from \cite{Leonard2015} are used. Since $\alpha^{\text{5s-6p+}}_{J}$ and $\alpha^{\text{c, cv}}$ are 4 orders of magnitude smaller than $\alpha^{\text{5s-5p}}_{F}$, their contribution is negligible in applications using far-off resonance dipole traps. In contrast, at the tune-out wavelength studied here, the polarizabilities of both Rb $D$ lines cancel, revealing these usually negligible components.
			
	Table \ref{tab:TheoryMagicWavelength} compares the theoretical prediction of the tune-out wavelength with our measurement. We first calculate a theoretical value for the tune-out wavelength $\lambda_{M, \text{5s-5p}}$, if only the $D$ line contributions $\alpha^{\text{5s-5p}}_{F}$ were present. Each further contribution, \textit{i.e.} the tensor polarizability, the higher transitions, and the core electrons leads to a correction of the tune-out wavelength toward higher wavelengths of in total $\Delta\lambda_M=\SI{4.749(87)}{\pico \meter}$.
	
	Using direct measurements of the reduced dipole matrix elements $d_{\nicefrac{1}{2}} = \Bra{5S_{\nicefrac{1}{2}}} d \Ket{5P_{\nicefrac{1}{2}}} $, and $d_{\nicefrac{3}{2}} = \Bra{5S_{\nicefrac{1}{2}}}d\Ket{5P_{\nicefrac{3}{2}}}$ in the calculation of the $D$ line contributions from \cite{Steck2010} yields an expected tune-out wavelength of $\SI{790.0228(57)}{\nano \meter}$. 
	Here, the uncertainty of $\SI{5.7}{\pico \meter}$ in the prediction does not allow for verifying the expected tune-out wavelength shift of $\Delta\lambda_M=\SI{4.749(87)}{\pico \meter}$.	
	For comparison, we take the more accurately measured ratio of both dipole matrix elements $R = \nicefrac{|d_{\nicefrac{3}{2}}|^2}{|d_{\nicefrac{1}{2}}|^2}$ from \cite{Leonard2015}, that has been gained from determining the tune-out wavelength of the $\Ket{F=2, m_F=2}$ state, and get $\SI{790.01850 (9)}{\nano \meter}$.  
	Our measurement is in agreement with the model within the $1\sigma$ uncertainties, confirming the non-negligible influence of higher transitions and the core-electron polarizability. 
	
	We emphasize that in the absence of the vector ac Stark shift for $m_F=0$, the apparent discrepancy with the tune-out wavelength value, obtained for Rb in $\Ket{F=2}$ of $\SI{790.0324}{\nano \meter}$ \cite{Leonard2015} results from different couplings to excited states rather than light polarization or magnetic field effects.
\subsection{$m_F=\pm 1$ State Measurement}
	Complementary information about the influence of the vector ac Stark shift is gained by investigating the lattice potential of $m_F=\pm 1$.
	The vector polarizability $\alpha^{\text{v}}_{F}(\lambda)$ is constant to the sub-percent level
	in the measured wavelength range, so the tune-out wavelengths for the $m_F=\pm 1$ states strongly depend on the polarization properties $A$ and $\theta_k$ of the light field.

	\begin{figure}[b]
		\begin{center}
			\includegraphics[width=0.5\textwidth]{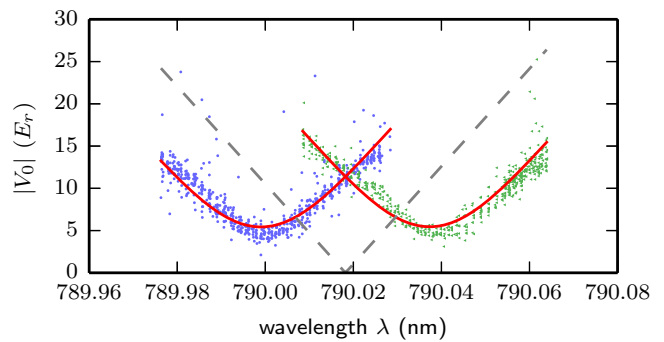}
		\end{center}
		\caption{Lattice potential for the $m_F=\pm 1$ states. Due to the vector Stark shift, the tune-out wavelength for the $m_F=-1$ state (circles, left) is shifted to a lower, and $m_F=+1$ (triangles, right) is shifted to a higher wavelength. As a reference, the $m_F = 0$ (dashed, center) potential is given. The solid line shows one single fit to both the $m_F=+1$ and the $m_F=-1$ state with our model, including a circular degree of polarization $A$ of the lattice (see eq. (\ref{eq:effectiveCircPolA})).
		}
		\label{fig:latticePolarizationState}
	\end{figure}

	The resulting vector light shift $V_0\propto A \cos{\theta_k} m_F \alpha^{\text{v}}_{F}$ yields a symmetric shift of the tune-out wavelengths with respect to $m_F=0$ according to the sign of the respective $m_F$ state.
	We perform the measurement analogously to $m_F=0$ for a quantization field of $B_z = \SI{50}{G}$, parallel to the $\hat{k}$ vector with an angle of $\theta_k = \SI{0.0(1)}{\degree}$. Therefore, the system is maximally sensitive to the degree of circular lattice polarization $A$.
	Figure \ref{fig:latticePolarizationState} shows the fitted lattice potentials for both states. The wavelength of minimal remaining lattice potential of $m_F=-1(+1)$ is shifted to a lower (higher) wavelength, indicating a small contribution of left hand circularly polarized ($\sigma^-$) lattice polarization. In addition, the lattice potential does not drop to zero as expected. We model this by a fluctuating degree of circular polarization during the KD pulse, effectively shaking the $v$-shape potential curve along the $\lambda$ axis and therefore smoothing out the minimum of the lattice potential. We assume a normally distributed probability of $A$ with a standard deviation of $\sigma_A$ around the expectation value $A_0$. This model yields an effective potential for Rb atoms of
	\begin{equation}
	\begin{aligned}
		|V_0| &= \frac{E_0^2}{4} \left[ \sqrt{\frac{8 {\sigma_A}^2}{\pi}} \text{e}^{- \frac{{\gamma}^2}{8 {\sigma_A}^2}} + \gamma \text{erf}{\left(\frac{\gamma}{\sqrt{8 {\sigma_A}^2}}\right)}	\right].   \\
	\end{aligned}
	\label{eq:effectiveCircPolA}
	\end{equation}
	Here, $\gamma = \alpha^{\text{s,t}} + A_0 \frac{m_F}{F}\alpha^{\text{v}}$ is the lattice potential in the absence of fluctuations in $A$ ($\sigma_A = 0$). We fit the model to the measured $m_F=\pm 1$ data, including the knowledge about the value of the magic wavelength $\lambda_M$ from the $m_F=0$ measurement, and using $\nicefrac{\partial V_0}{\partial \lambda}$, $A_0$ and $\sigma_A$ as only free parameters. We obtain $A_0 = \SI{-7.80(4)e-3}{}$, which corresponds to an angle of circular polarization of $\theta_0 = \SI{-0.223(1)}{\degree}$,  and a fluctuation of $\sigma_A = \SI{4.78(9)e-3}{}$ and $\sigma_{\theta_0}= \SI{0.137(8)}{\degree}$, respectively.
	In our setup both lattice arms are linearized before entering the vacuum chamber by a combination of a polarizing beam splitting cube (PBC), a polarization maintaining optical fiber, a half-wave plate, and another PBC.
	Therefore, we attribute the admixture of circular polarization components to the birefringence of the vacuum chamber windows \cite{Solmeyer2011, Steffen2013, Zhu2013}, which however does not explain the comparably large and high-frequency fluctuations.
	
	The results from the $m_F=\pm 1$ measurements are of major importance for a possible application of the lattice for species-selective experiments, since even for $A_0=0$ and the magnetic field orientation chosen here, the fluctuations lead to a non-vanishing lattice potential for $m_F = \pm 1$ of more than $\SI{5}{E_r}$. For comparison, in the $m_F=0$ state the residual lattice potential dropped to $\SI{1.1}{E_r}$, when averaging in a wavelength range of $\pm\SI{1}{\pm}$ around the tune-out wavelength.
	\subsection{Magnetic Field Dependence}
	\begin{figure}[tb]
		\begin{center}
			\includegraphics[width=0.5\textwidth]{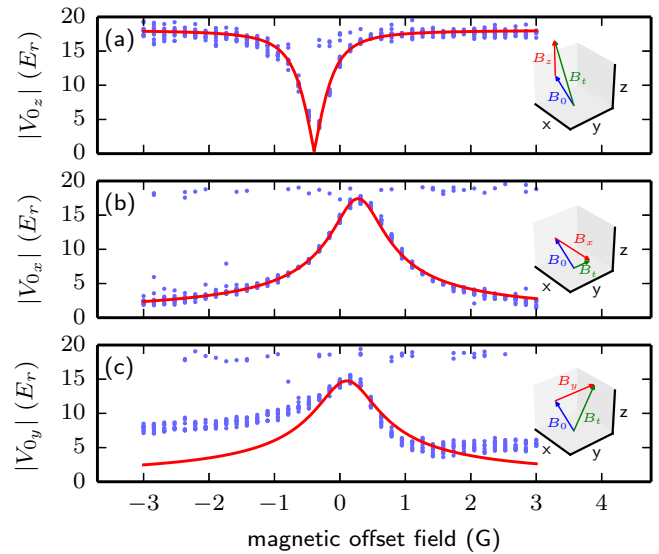}
		\end{center}
		\caption{Measurement of the vector Stark shift. (a) An offset field $B_z$ along the $z$ axis, parallel to the lattice $\vec{k}$ vector is applied. For $B_z=-B_{0,z}$, the quantization axis along the total magnetic field $\vec{B}_t$ is perpendicular to $\vec{k}$, minimizing the vector light shift. The minimum at $B_{0,z}$ is fitted (solid line) according to our model (eq. \ref{eq:thetaKMagnetic}). (b) For an offset field $B_x$ along the $x$ axis, the projection of $\vec{B}_t$ is maximized for $B_x=-B_{0, x}$. Using $B_{0,z}$ from (a), we get $B_{0,x}$ and $B_{0,y}$ from the fit (solid line). (c) Applying an offset $B_y$ along the $y$ axis, the vector Stark shift is maximized for $B_y = -B_{0,y}$ as well. The solid line shows the model with fit results from (a) and (b). The lattice wavelength was held constant at the tune-out wavelength $\SI{790.0182(2)}{\nano \meter}$ for all measurements. For each magnetic field, roughly 20 data points were taken.}
		\label{fig:MagneticFieldMeasurement}
	\end{figure}
	After studying the dependence of the vector ac Stark shift on the light polarization, we investigate the influence of the magnetic field orientation on the lattice potential. The vector Stark shift is proportional to the projection $\cos{\theta_k}={\vec{k} \cdot \vec{B}_t} / (|\vec{k}| |\vec{B}_t|)$ of the lattice $\vec{k}$ vector to the total magnetic field $\vec{B}_t$. We superpose the magnetic background $\vec{B}_0$ with a known offset field $\vec{B} = (B_x, B_y, B_z)$, and thereby vary $\vec{B}_t$. With $\vec{k}$ aligned parallel to the $z$ axis, the projection writes
	\begin{equation}
		\cos{\theta_k} = \frac{B_{z, t}}{\sqrt{B_{x, t}^2 + B_{y, t}^2 + B_{z, t}^2}}
	\label{eq:thetaKMagnetic}
	\end{equation}
	with the spatial components of the total magnetic field $B_{i, t}=B_{0, i} + B_i$ in each direction $i$. Important cases are $B_z=-B_{0,z}$, where the projection is vanishing as well as $B_x=-B_{0,x}$ and $B_y=-B_{0,y}$, where $\cos{\theta_k}$ is maximized.	
	We measure the vector ac Stark shift for varying offset along one direction, while keeping the fields in both remaining directions at zero.
	The lattice wavelength is set to the tune-out wavelength of $m_F=0$, allowing for maximum sensitivity to the vector ac Stark shift.
	
	Figure \ref{fig:MagneticFieldMeasurement} shows the measured lattice potential for Rb in the $m_F=+1$ state. From the $B_z$, and the $B_x$ variation measurement, we obtain a background field of $\vec{B}_0 = (0.28, 0.11, -0.39)\SI{}{G}$. 
	For comparison, using a Hall probe, we measure a magnetic background field of $\vec{B}_{0,H}=(0.25(1), 0.11(1), -0.20(3))\SI{}{G}$ in the lab.
	While $B_{0,x}$, and $B_{0,y}$ agree fairly well with the independent measurement, a discrepancy in $B_{0,z}$ occurs, which we attribute to an additional magnetic background from our setup due to a residual field from the homogeneous field coils. Using the fitted background field, the lattice potential maximum in the $B_y$ variation is reproduced. We suspect the asymmetric wavelength to result from an unwanted magnetic field component of the $B_y$ coils along the $z$ direction. To increase the accuracy of the background field measurement, optical magnetometry \cite{Budker2007} or magnetic field imaging \cite{Koschorreck2011} techniques could be applied.

	Combining our results of the lattice polarization measurement and the magnetic background field $\vec{B}_0$, we gained a valuable understanding of the factors $A$ and $\cos{\theta_k}$, that determine the influence of the vector Stark shift on the lattice potential \textit{in situ}. In particular, by applying a strong offset field along the $x$ axis, we can reduce the influence of the polarization fluctuations, reaching equally vanishing trapping potentials for all $m_F = 0, \pm 1$ states at the tune-out wavelength, as indicated in the measurement, shown in figure \ref{fig:MagneticFieldMeasurement} (b).
	
\section{Conclusions}
	We have presented an experiment to measure the ac Stark shift around the tune-out wavelength of Rb in the hyperfine ground state manifold $\Ket{F=1, m_F=0,\pm1}$ at $\SI{790}{\nano \meter}$. At the tune-out wavelength, ac Stark shifts from higher transitions and core electrons, as well as vector and tensor polarizabilities are resolved, that are orders of magnitude smaller than the dominant scalar polarizabilities from the $D_1$ and $D_2$ line. In addition, by separating the magnetic $m_F=0, \pm1$ Zeeman states, we exclude the influence of the vector ac Stark shift on demand. Our measurements feature a Kapitza-Dirac scattering technique, combined with an improved absorption image processing, the absence of additional trapping light fields, and a magnetically controlled environment.
	
	When measuring the $m_F=0$ state, we exclude the influence of the vector ac Stark shift on the tune-out wavelength. Our value of $\SI{790.01858 (23)}{\nano \meter}$ provides a 10-fold accuracy improvement compared to the previous measurement \cite{Lamporesi2010} in the same atomic state, and therefore allows for resolving the influence of transitions to higher principle quantum numbers, core electron and core-valence electron contributions to the scalar polarizability $\alpha^{\text{s}}$ of Rb. 
	This confirms a recent measurement of these contributions \cite{Leonard2015} in a complementary system.

	The vector ac Stark shift is included into the system when measuring the lattice potential of the $m_F=-1$ and $m_F=+1$ states. From the shift of the tune-out wavelength with respect to $m_F=0$, we evaluate the degree of circular polarization $A_0 = \SI{-7.80(4)e-3}{}$ of the optical lattice, and its fluctuation  $\sigma_A = \SI{4.78(9)e-3}{}$ \textit{in situ}. In addition, by exploiting the dependence of the vector light shift on the quantization field orientation, we have determined the magnetic background field.\\
	
	Besides probing the atomic level structure beyond common approximations, we apply the tune-out wavelength in our Cs-Rb mixed-species experiment. For our species-selective trapping application, we reach a trap potential selectivity $\nicefrac{V_\text{Cs}}{V_\text{Rb}}$ exceeding $\SI{1.8e3}{}$ for the $m_F=\pm1$ states and more than $\SI{3.3e3}{}$ in the case of $m_F=0$. In a lattice physics scenario with lattice depth for Cs in the order of $\SI{25}{E_{r,\text{Cs}}}$ \cite{Bloch2005}, for Rb this would cause a remaining trap potential of $\SI{9.1e-3}{E_{r}}$, and a photon scattering rate of $\SI{5.7}{\Hz}$, where $E_{r,\text{Cs}}=\nicefrac{\left(\hbar k\right)^2}{2m_{\text{Cs}}}$ is the recoil energy for Cs.
	The species-selective lattice will allow for the study of non-equilibrium interaction effects, such as polaron transport \cite{Bruderer2008}, coherence properties of Cs in the Rb bath \cite{Klein2005}, and Bloch oscillations \cite{Grusdt2014}. Moreover, a full understanding of all relevant ac Stark shift contributions to the Rb potential enables us to engineer $m_F$ state-dependent trapping schemes with variable selectivity and tunable species overlap.
\section{Acknowledgments}
	The project was financially supported partially by the European Union via the ERC Starting Grant 278208 and partially by the DFG via SFB/TR49. D.M. is a recipient of a DFG-fellowship through the Excellence Initiative by the Graduate School Materials Science in Mainz (GSC 266), F.S. acknowledges funding by Studienstiftung des deutschen Volkes, and T.L. acknowledges funding from Carl-Zeiss Stiftung.

\bibliographystyle{apsrev4-1}
\bibliography{manuscript}

\end{document}

%% file: meta/packages.tex
\usepackage[utf8x]{inputenc}
\usepackage{color}
\usepackage{bbm} 

\usepackage{nicefrac}
\usepackage{amsfonts,amsmath,amssymb,stmaryrd}
\usepackage{braket}

 \usepackage[autostyle=true]{csquotes}

\usepackage{tabularx}
\usepackage{multirow} 
\usepackage{hhline}
\usepackage{graphicx}
\usepackage{subfigure}  
\usepackage{bbm} 
\usepackage[pdftex]{epsfig}
\usepackage{mathrsfs}
\usepackage{verbatim}
\usepackage{centernot}
\usepackage{ulem}
\usepackage{array}
\usepackage{cancel}
\usepackage{ifthen}
\usepackage{bm}	
\usepackage{siunitx}
\usepackage{todonotes}
\InputIfFileExists{gitinfo-latex.inc}{}{}

\usepackage[acronym,nomain]{glossaries}

%% file: meta/authors.tex
\author{Felix Schmidt}
\affiliation{Department of Physics and Research Center OPTIMAS, University of Kaiserslautern, Germany}
\affiliation{Graduate School Materials Science in Mainz, Gottlieb-Daimler-Strasse 47, 67663 Kaiserslautern, Germany}

\author{Daniel Mayer}
\affiliation{Department of Physics and Research Center OPTIMAS, University of Kaiserslautern, Germany}
\affiliation{Graduate School Materials Science in Mainz, Gottlieb-Daimler-Strasse 47, 67663 Kaiserslautern, Germany}

\author{Michael Hohmann}
\affiliation{Department of Physics and Research Center OPTIMAS, University of Kaiserslautern, Germany}

\author{Tobias Lausch}
\affiliation{Department of Physics and Research Center OPTIMAS, University of Kaiserslautern, Germany}

\author{Farina Kindermann}
\affiliation{Department of Physics and Research Center OPTIMAS, University of Kaiserslautern, Germany}

\author{Artur Widera}
\affiliation{Department of Physics and Research Center OPTIMAS, University of Kaiserslautern, Germany}
\affiliation{Graduate School Materials Science in Mainz, Gottlieb-Daimler-Strasse 47, 67663 Kaiserslautern, Germany}